\documentclass[showpacs,aps,twocolumn]{revtex4}
\usepackage{amsmath}
\usepackage{amsfonts}
\usepackage{amssymb}
\usepackage{amsthm}
\usepackage{graphicx}
\usepackage{color}

\begin{document}


\title{Giant dynamical Zeeman split in inverse spin valves}

\author{X. R. Wang}
\affiliation{ Department of Physics, The Hong Kong University 
of Science and Technology, Clear Water Bay, Hong Kong, China}
\date{\today}

\begin{abstract}
The inversion of a spin valve device is proposed.
Opposite to a conventional spin valve of a non-magnetic spacer 
sandwiched between two ferromagnetic metals, an inverse spin
valve is a ferromagnet sandwiched between two non-magnetic metals.
It is predicted that, under a bias, the chemical potentials
of spin-up and spin-down electrons in the metals split at
metal-ferromagnet interfaces, a dynamical Zeeman effect.
This split is of the order of an applied bias.
Thus, there should be no problem of generating an $eV$ split
that is not possible to be realized on the earth by the usual
Zeeman effect.
\end{abstract}
\keywords{Spin valve, Zeeman effect, Spintronics, Magnetism}
\pacs{} \maketitle

Spintronics becomes an emergent subfield in condensed matter 
physics since the discovery of giant magnetoresistance 
in 1988 by Fert\cite{Fert} and Grunberg\cite{Grun}. 
Spintronics is about the control and manipulation of both 
electron charge and electron spin. Many interesting phenomena 
are related to the interplay between electron spin and its 
charge degrees of freedom. For example, electron transport 
can be manipulated by the magnetization configurations. 
This is the basis of giant magnetoresistance\cite
{Fert,Grun,Parkin} and tunneling magnetoresistance\cite
{Miyazaki,Moodera} phenomena and devices. 
Its inverse effect, known as spin-transfer torque 
(STT)\cite{Slon,Berger,xrw}, was also discovered. 
The STT opens a new way to manipulate magnetization other 
than a magnetic field\cite{xrw1}, which has been much of 
recent focus in the field due to its potential applications 
in information storage industry. 

In this letter, an inverse spin valve structure of a ferromagnet 
sandwiched between two non-magnetic metals is studied when 
a bias is applied to the device. Due to the spin-dependent 
electron transport of the structure, the chemical potentials 
of spin-up (SU) and spin-down (SD) electrons at the 
metal-ferromagnet interfaces split by the magnitude of an 
applied bias. We term this split dynamical Zeeman effect. 
This split can be of the order of $eV$ if proper materials are 
used. To introduce a similar split by the usual Zeeman effect, 
a magnetic field unrealizable on the earth is required. 
Thus, a giant dynamical Zeeman effect is predicted. 

A conventional spin valve is a layered structure of a 
non-magnetic spacer sandwiched between two ferromagnetic metals. 
The spin valve is called a giant magnetoresistance device 
if the spacer is a normal metal while it is a tunneling 
magnetoresistance device for an insulator spacer. 
The electron transport of a spin valve depends on the relative 
polarities of the two magnets. An inverse spin valve is also a 
layered structure with a ferromagnet sandwiched between two 
non-magnetic metals as illustrated in Fig. 1a. $M1$ and $M2$ 
are two normal metals with identical SU and SD electron density 
of states (DOS) as depicted schematically in the left and right 
diagrams of Fig. 1b. To simplify our analysis, a half-metal (HM) 
spacer is considered first so that only electrons of one type of 
spins (spin-up) can pass through it. The DOSs of SU and SD 
electrons for a half-metal is illustrated by the middle diagram 
of Fig. 1b. Under a bias $V$, SU electrons in $M1$ flow into the 
empty SU electron states in $M2$ via the empty SU electron states 
in the half-metal, shown pictorially by the curved arrows in Fig. 
1b. The flow of the electrons creates chemical potential 
differences ($\Delta\mu_i$ in Fig. 1b) between SU and SD electrons 
in both $M1$ and $M2$ near the metal-ferromagnet interfaces. 
\begin{figure}[htbp]
 \begin{center}
\includegraphics[width=7.cm, height=4.cm]{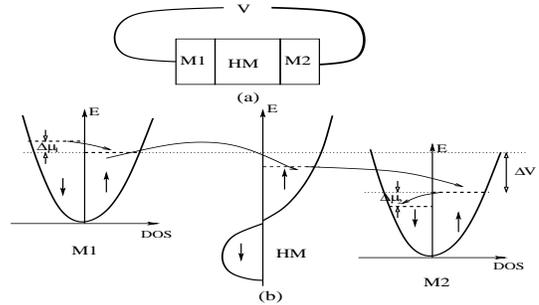}
 \end{center}
\caption{\label{fig1} Schematic illustration of the inverse 
spin valve (a) and relative chemical potentials of spin-up and 
spin-down electrons in nonmagnetic metals and half-metal (b). 
The curved arrows indicate the electron flow. $\Delta\mu_1$ 
and the $\Delta\mu_2$ are the chemical potential splits 
between spin-up and spin-down electrons at the left and right 
metal-ferromagnet interfaces. $\Delta V$ is the effective bias 
on the half-metal.}
\end{figure}

In order to understand why the chemical potentials of SU and SD
electrons at the $M1-HM$ and $M2-HM$ interfaces split under a
bias, let us consider an extreme case in which spin relaxation
time in both $M1$ and $M2$ is infinite long (no spin relaxation
so that SU and SD electrons are isolated from each other).
Since electron transport in a nanostructure depends on how a
bias is applied\cite{sdwang}, we assume, to be precise and
without losing generality, that the electron chemical
potential in $M1$ is initially moved up by $eV$ while that of
$M2$ is kept unchanged. Initially, SU electrons flow into the
empty SU electron states in $M2$ because the half-metal
prevent SD electrons from flowing. The same amount electrons
will be pumped back from $M2$ to $M1$ by a battery to keep the
electron neutrality in $M1$ and $M2$. However, a battery does
not distinguish electron spin, and as a result it will pump
equal amount of SU and SD electrons.
In other words, SU electrons flow out of $M1$ and into $M2$ via
the half-metallic spacer. In the meanwhile, an equal amount of
electrons with half of them being in the SU state and the other
half in the SD state are drawn out of $M2$ and are supplied
into $M1$ by the battery. As a result, $M1$ accumulates
more SD electrons, and $M2$ accumulates more SU electrons.
Thus, the chemical potential of the SD electrons will be higher
than that of SU electrons in $M1$. Vice versa, the chemical
potential of the SU electrons will be higher than that of SD
electrons in $M2$. The chemical potential splits keep increasing
until the chemical potential of SU electrons in $M1$ equals
that in $M2$. At this point, the steady state with no current
in the circuit is reached and chemical potential splits in $M1$
and $M2$ are established with $\Delta\mu_1+\Delta\mu_2=eV$.

The spin relaxation\cite{t1t2} always exists in a material,
and the magnitude of the chemical potential split for SU and
SD electrons depends sensitively on the spin relaxation.
This sensitivity can be seen from another extreme case that the
spin flipping is so fast that SU and SD electrons can be
converted into each other any time at no cost (meaning SU and
SD electrons are at equilibrium with respect to each others
at all times in $M1$ and $M2$). Thus, it is impossible to
create any chemical potential difference for SU and SD
electrons in $M1$ and $M2$. The reality, of course,
is somewhere in between (the two extreme cases).
The spin relaxation time of a real material is finite, and it
depends on the strength of spin-orbital coupling, hyperfine
interaction and other interactions that cause spin flipping.
In order to include the spin relaxation time quantitatively,
consider an ideal model at zero temperature. Assume the spin
flip occurs only near the interfaces within a width of spin
diffusion length $\xi_1$ in $M1$ and $\xi_2$ in $M2$.
This is justified because both SU and SD electrons in the rest
parts of the circuit (other than the half-metal) should have
same chemical potential. Thus number of electrons flipped
from up-spin state to down-spin state is the same as that from
down-spin to up-spin. Furthermore, let us assume the resistance
of the half-metal is $R$ (for SU electrons, the resistance for
SD electrons is infinity due to the half-metallic nature of the
middle spacer).
\begin{figure}[htbp]
 \begin{center}
\includegraphics[width=7.cm, height=4.cm]{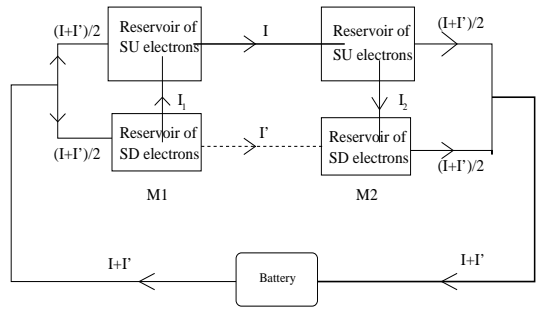}
 \end{center}
\caption{\label{fig2} Schematic illustration of current flow 
from and into spin-up and spin-down states in $M1$ and $M2$. 
At the steady state, current flowing into any reservoir should 
be equal to those flowing out. }
\end{figure}

The equations of the motion of the device can be obtained by
considering electron flow diagram of Fig. 2. There are two
reservoirs in $M1$ and $M2$ each. One is for SU electrons, and
the other is for SD electrons (denoted by rectangular boxes).
SU electrons in $M1$ can flow into SU electron reservoir of $M2$.
The current $I$ depends on resistance $R$ and chemical potential
difference $\Delta V$ (Fig. 1b) of SU electrons in $M1$ and $M2$,
\begin{equation}\label{current}
I=\frac{\Delta V}{R}. 
\end{equation}
If one neglects the direct tunneling of SD electron from
$M1$ to $M2$ through the half-metal ($I'=0$ in Fig. 2), SD
electrons can only go to $M2$ by first flipping their spins
and converting themselves into SU electrons. Let $\tau_1$
and $\tau_2$ be the spin flipping time (spin-relaxation time
$T_1$\cite{t1t2}) in $M1$ and $M2$, corresponding to the
flipping rate of $1/\tau_1$ and $1/\tau_2$, respectively.
Due to the conversion of SD electrons to SU electrons that
is the product of the excess SD electrons $n_1 \Delta\mu_1 
\xi_1 A$ and the single electron flipping rate, the current
from SD reservoir to SU reservoir in $M1$ is
\begin{equation}\label{current1}
I_1=\frac{n_1\Delta \mu_1e\xi_1 A }{\tau_1},
\end{equation}
where $n_1$ is the density of states of SD electrons in
$M1$ at the Fermi level. $A$ is the cross section of $M1$.
Similarly, the current due to the conversion of SU electrons
to SD electrons in $M2$ is
\begin{equation}\label{current2}
I_2=\frac{n_2\Delta \mu_2e\xi_2 A }{\tau_2}.
\end{equation}
At the steady state, there is no net electron build up anywhere
in the circuit. Since the current through the battery is
unpolarized, spin-up electrons will be mixed up with spin-down
electrons and be pumped by the battery from $M2$ into $M1$.
Thus, the SU electrons make up half of the current while
other half is made up from the SD electrons, and the balancing
conditions and external constraint require
\begin{align}\label{steady}
&I_1=I/2 ,\nonumber\\
&I_2=I/2 , \\
&\Delta\mu_1/e+\Delta\mu_2/e+\Delta V=V. \nonumber 
\end{align}
Solving Eqs. \eqref{current}, \eqref{current1},\eqref{current2},
and \eqref{steady}, the dynamical Zeeman split $\Delta\mu_1$ 
and $\Delta\mu_2$ are 
\begin{align}\label{split}
&\Delta\mu_1=\frac{(eV)\tau_1/(n_1e^2\xi_1A)}
{2R+\tau_1/(n_1e^2\xi_1A)+\tau_2/(n_2e^2\xi_2A)},\nonumber \\ 
&\Delta\mu_2=\frac{(eV)\tau_2/(n_2e^2\xi_2A)}
{2R+\tau_1/(n_1e^2\xi_1A)+\tau_2/(n_2e^2\xi_2A)}.
\end{align}
It is interesting to see that the largest dynamical Zeeman 
split occur at $R=0$, a short circuit for spin-up (SU) 
electrons! In this particular case, one, of course, needs 
to use metals with proper material parameters such that 
current density is not too large to cause the metal breaking 
down by heating. Further discussions on this issue are given 
soon. 

The half-metal may be replaced by a usual ferromagnetic metal. 
In this case, SD electrons in $M1$ can also flow directly 
into $M2$, contributing an extra current $I'$ to the circuit  
\begin{equation}\label{currentSD}
I'=\frac{V}{R'}, 
\end{equation}
where $R'$ is the resistance of ferromagnet for SD electrons.  
Without losing generality, the minority carriers of the 
ferromagnet are assumed to be the SD electrons, and $R'$ is 
also assumed to be larger than $R$. 
The chemical potential difference of SD electrons in $M1$ and 
$M2$ equals $V$ as it is shown in Fig. 1b. Eq. \eqref{steady} 
should be modified accordingly as  
\begin{align}\label{steady1}
&I_1=(I-I')/2 ,\nonumber\\
&I_2=(I-I')/2 , \\
&\Delta\mu_1/e+\Delta\mu_2/e+\Delta V=V. \nonumber 
\end{align}
This set of equations with non-zero $I'$ can be solved, and 
the dynamical Zeeman split $\Delta\mu_1$ and $\Delta\mu_2$, 
in comparison with that of half-metal case, are reduced 
by a factor of $(1-R/R')$ 
\begin{align}\label{split1}
&\Delta\mu_1=\frac{eV(1-R/R')\tau_1/(n_1e^2\xi_1A)}
{2R+\tau_1/(n_1e^2\xi_1A)+\tau_2/(n_2e^2\xi_2A)},\nonumber \\ 
&\Delta\mu_2=\frac{eV(1-R/R')\tau_2/(n_2e^2\xi_2A)}
{2R+\tau_1/(n_1e^2\xi_1A)+\tau_2/(n_2e^2\xi_2A)}.
\end{align}
One should not be surprised about this reduction from our early 
explanations of the origin of this split. Obviously, previous 
results Eq. \eqref{split} are recovered when $R'=\infty$. 
Also, there are no chemical potential splits in $M1$ and $M2$ 
when the spacer is non-magnetic ($R=R'$). 
Therefore, a good half-metal (good conductor for the majority 
carriers and good insulator for the minority carriers) should 
be used if one wants to maximize the split. In the following 
discussion, $R'=\infty$ case is considered only. 

The conventional way of introducing an energy split for 
SU and SD electrons is through the Zeeman effect. 
Due to the small value of Bohr magneton, an order of 
$10T$ field can only induce about $1meV$ energy split. 
However, the dynamical Zeeman split predicted here could 
easily be of the order of $1eV$, a truly giant Zeeman effect. 
Compare with the static Zeeman effect, this split is 
equivalent to a field in the order of $10^4T$, an impossible 
magnetic field on the earth! 

It is interesting to notice that a large spin-dependent 
chemical potential difference means a large dynamical magnet. 
For $\Delta\mu_i=1eV$ and a typical electron density of 
states of $10^{22}eV^{-1}cm^{-3}$ for a metal, the dynamical 
magnetization is about $10^5A/m$ which is comparable with 
many magnets. Thus one can use MOKE (magneto-optical 
Kerr effect) to `see' the dynamical magnetization. 
Upon verification of this dynamical magnetism, it should be 
interesting to explore the potential applications of this 
electric-field controlled magnet. Another possible 
application of the predicted phenomena is polarized 
electron/light source\cite{xrw2}. Since electrons of one 
type of spins occupy higher energy levels than those of the 
opposite spin, the predicted effect can be used as a 
polarized electron source, or light source when the electrons 
flip their spins and emitted well-defined polarized photons. 
Thus, the phenomenon can be used to make tunable light 
emitting diode or laser in very wide frequency range. 
The large chemical potential difference may also be used 
to enhance electron magnetic resonance signal. 
The electron magnetic resonance is useful in probing material 
properties in various technologies with wide applications, 
including imaging in information processing. 

It should be pointed out that we have not considered the 
spatial distribution of the chemical potential split. 
Our results may be modified quantitatively, but not 
qualitatively, when the detailed distribution is taken 
into account. This is because a factor characterized the 
distribution should be added to equations (3) and (4).
It should also be noted that the giant dynamical Zeeman 
split reported here is different from the split in the DOS 
of a ferromagnet. In the usual ferromagnets, electron 
DOS is spin dependent, but the chemical potential of both 
SU and SD electrons are the same at the equilibrium. 
However, the dynamical Zeeman split predicted here occurs 
inside a non-magnetic metal where the electron DOS is 
spin-independent, but the spin-up and spin-down electrons 
fill their DOSs to different levels, leading to a chemical 
potential difference.

In conclusion, we propose an inverse spin valve structure 
consisting of a ferromagnet sandwiched between two normal metals. 
Under a bias, we predicted a giant dynamical Zeeman split 
of the chemical potential for spin-up electrons and 
spin-down electrons at metal-ferromagnet interfaces. 
This prediction is yet to be confirmed by experiments. 

The author would like to thank Prof. Yongli Gao and Prof. John 
Xiao for the useful discussions. This work is supported by UGC, 
Hong Kong, through CERG grant (\# 603007 and SBI07/08.SC09).


\begin{thebibliography}{}
\bibitem{Fert}  M.N. Baibich, J.M. Broto, A. Fert, F.N. Van 
Dau, F. Petroff, P. Etienne, G. Creuzet, A. Friederich, 
and J. Chazelas, Phys. Rev. Lett. {\bf 61}, 2427 (1988). 
\bibitem{Grun} G. Binach, P. Grunberg, F. Saurenbach, 
and W. Zinn, Phys. Rev. B {\bf 39}, 4828 (1989). 
\bibitem{Parkin} R.F.C. Farrow, C.H. Lee, and S.S.P. Parkin, 
IBM J. Res. Dev. {bf 34}, 903 (1990). 
\bibitem{Miyazaki} T. Miyazaki and N. Tezuka, J. Magn. Magn. 
Mater. {\bf 139}, L231 (1995).   
\bibitem{Moodera} J.S. Moodera et al., Phys. Rev. Lett. 
{\bf 74}, 3273 (1995).
\bibitem{Slon} J.C. Slonczewski, J. Magn. Magn. Mater. {\bf 159},
L1-L7 (1996).
\bibitem{Berger} L. Berger, Phys. Rev. B {\bf 54}, 9353 (1996).
\bibitem{xrw} X.R. Wang, and Z.Z. Sun, Phys. Rev. Lett.
{\bf 98}, 077201 (2007).
\bibitem{xrw1} Z.Z. Sun, and X.R. Wang, Phys. Rev. Lett. 
{\bf 97}, 077205 (2006); Phys. Rev. B {\bf 71}, 174430 (2005); 
{\it ibid} {\bf 73}, 092416 (2006); {\it ibid} {\bf 73}, (2007);
T. Moriyama, R. Cao, J.Q. Xiao, J. Lu, X.R. Wang, Q. Wen, 
and H.W. Zhang, Appl. Phys. Lett. {\bf 90}, 152503 (2007). 
\bibitem{sdwang} S.D. Wang,  Z.Z. Sun, N. Cue, H.Q. Xu, and 
X.R. Wang, Phys. Rev. B {\bf 65}, 125307 (2002).
\bibitem{t1t2} X.R. Wang, Y.S. Zheng, and S. Yin, Phys. Rev. B 
{\bf 72}, R121303 (2005).
\bibitem{xrw2} X.R. Wang, unpublished.
\end{thebibliography}
\end{document}